\documentclass[aip, amsmath,amssymb, reprint,]{revtex4-1}

\usepackage{graphicx}
\usepackage{dcolumn}
\usepackage{bm}
\usepackage[utf8]{inputenc}
\usepackage{multirow}
\usepackage[T1]{fontenc}
\usepackage{mathptmx}
\usepackage{etoolbox}
\usepackage{graphicx}

\makeatletter
\def\@email#1#2{
 \endgroup
 \patchcmd{\titleblock@produce}
  {\frontmatter@RRAPformat}
  {\frontmatter@RRAPformat{\produce@RRAP{*#1\href{mailto:#2}{#2}}}\frontmatter@RRAPformat}  {}{}}
\makeatother
\begin{document}

\preprint{AIP/123-QED}

\title{A way to identify whether a DFT gap is from right reasons or error cancellations: The case of copper chalcogenides} 

\author{Jiale Shen}
\affiliation{Key Lab of advanced optoelectronic quantum architecture and measurement (MOE), and Advanced Research Institute of Multidisciplinary Science, Beijing Institute of Technology, Beijing 100081, China}
\author{Haitao Liu}
\affiliation {Institute of Applied Physics and Computational Mathematics, Beijing 100088, China}
\affiliation {National Key Laboratory of Computational Physics, Beijing 100088, China}
\author{Yuanchang Li}
\email{yuancli@bit.edu.cn}
\affiliation{Key Lab of advanced optoelectronic quantum architecture and measurement (MOE), and Advanced Research Institute of Multidisciplinary Science, Beijing Institute of Technology, Beijing 100081, China}

\date{\today}

\begin{abstract}
Gap opening remains elusive in copper chalcogenides (Cu$_{2}X$, $X$ = S, Se and Te), not least because Hubbard + $U$, hybrid functional and ${GW}$ methods have also failed. In this work, we elucidate that their failure originates from a severe underestimation of the 4$s$-3$d$ orbital splitting of the Cu atom, which leads to a band-order inversion in the presence of an anionic crystal field. As a result, the Fermi energy is pinned due to symmetry, yielding an invariant zero gap. Utilizing the hybrid pseudopotentials to correct the underestimation on the atomic side opens up gaps of experimental magnitude in Cu$_{2}X$, suggesting their predominantly electronic nature. Our work not only clarifies the debate about the Cu$_{2}X$ gap, but also provides a way to identify which of the different methods really captures the physical essence and which is the result of error cancellation.
\end{abstract}

\pacs{}

\maketitle 

\section{Introduction}

Understanding the nature of the gap and accurately calculating its size have been one of the central themes of modern electronic structure theory, especially in the study of transition-metal compounds with exotic physics and promising applications\cite{Rao,Imada,Tokura,Zaanen}. Unfortunately, this remains a formidable challenge for today's widely used calculation methods. For example, density-functional theory (DFT) systematically underestimates gaps and can even incorrectly predict semiconductors as metals\cite{Chan,LiW}. To address this problem, methods such as Hubbard + $U$, hybrid functionals and ${GW}$ have been developed based on DFT, which typically reproduce the experimental gap under certain conditions\cite{Tran,Anisimov,Heyd,Hybertsen}. However, there is a lack of ways to identify which truly capture the physical essence and which are the result of error cancellation.

Copper chalcogenides (Cu$_{2}X$, $X$ = S, Se and Te) are a class of low-cost and environmentally friendly materials for thermoelectric and photovoltaic applications\cite{Coughlan,x-zt-2012}. Their interesting phase shares a common structure in which the $X$ atoms form an ordered lattice skeleton while the Cu atoms are randomly distributed on tetrahedral interstitial sites formed by the $X$ atoms. It has been known for a long time that Cu$_2X$ has an optical gap in the range 1.08$\sim$1.23 eV\cite{s-gap-1965, se-gap-1966, te-gap-1990}, yet theoretical study of its electronic structure has been hindered by the complexity of its crystal structure. Until 2007, Lukashev et al. propose a highly symmetric antifluorite model that has since been widely adopted\cite{s-lda-prb-2007, se-jpcm-2013, te-prb-2022}. Using this perfect crystal structure, their LDA calculation yields a zero gap characterized by the intersection of the conduction and valence bands at the $\Gamma$ point. This is not surprising, since it had become generally known by then that DFT underestimates the gap, especially given the presence of the transition-metal Cu. What is really surprising is that the Hubbard + $U$, hybrid functionals and quasiparticle self-consistent ${GW}$ calculations also feature the same zero gap\cite{s-lda-prb-2007,se-jpcm-2013,x-mbj-jcp-2014}. Remarkably, the self-consistent ${GW}$ succeeds in solving gap underestimation of analogous monovalent copper halides and oxides\cite{Schilfgaarde,Bruneval} that have long been notorious for DFT and its derivatives. In view of the failure of ${GW}$, Lukashev et al. explore the possibility of opening the gap through a "so-called structural mechanism" by considering the deviation of the realistic structure from the antifluorite model\cite{s-lda-prb-2007}. However, their calculations show that this would only yield about half of the experimental gap. Since then, there has been much debate as to whether the gap of Cu$_{2}X$ is opened by an electronic or structural mechanism. In 2012, Xu et al.\cite{Xu} use the HSE to compute much more realistic structures, obtaining gaps that are almost identical to the experiments. This seems to indicate that the structural mechanism is indeed at work. However, in 2014, Zhang et al.\cite{x-mbj-jcp-2014} find that the mBJ + $U$ calculations can open up 0.5$\sim$1.2 eV gap for the perfect antifluorite structure, reigniting the controversy.

Does the gap of Cu$_{2}X$ arise from an electronic mechanism or from a structural mechanism due to the over-simplified antifluorite model? Is there some commonality in gap underestimation for monovalent copper chalcogenides, halides and oxides? Despite years of attention, these mysteries remain unknown.

\begin{figure*}[htbp]
\includegraphics[width=1.9\columnwidth]{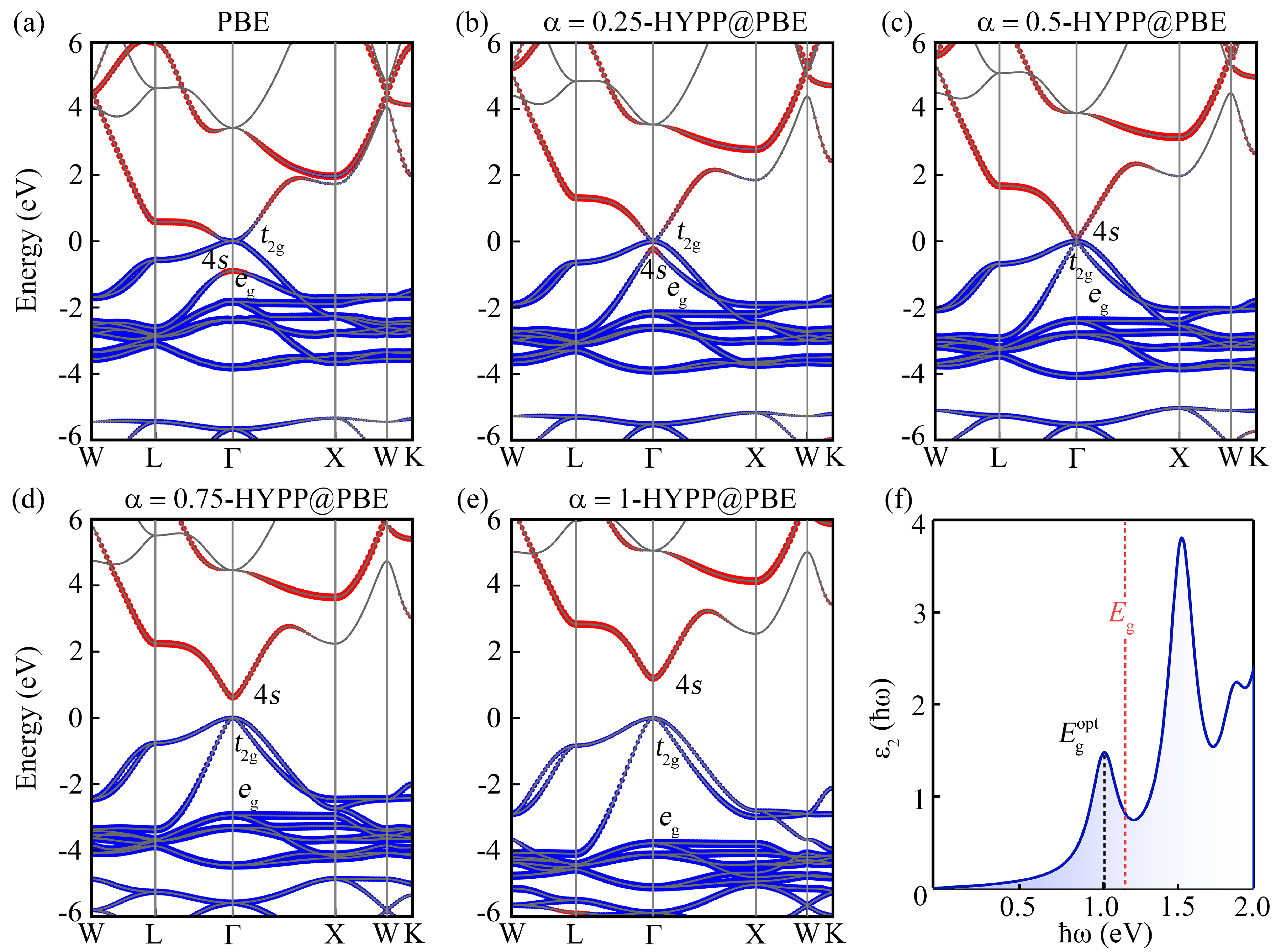}
\caption{\label{fig:fig1} Band structures of Cu$_2$S by (a) standard PBE, (b)-(e) HYPP@PBE at different $\alpha$. Red and blue dots represent the contributions from Cu 4$s$ and 3$d$ orbitals, respectively. Three characteristic bands are marked as $t_{\rm 2g}$, 4$s$ and $e_{\rm g}$ at the $\Gamma$ point. The valence band maximum is set as energy zero. (f) Imaginary part of the BSE dielectric function calculated on top of $\alpha$ = 1-HYPP@PBE results. Black and red dashed lines denote optical and fundamental gaps, respectively.}
\end{figure*}

Perhaps it is both electronic and structural mechanisms that are responsible for the gap. Under such circumstances, ones must quantitatively compare the respective sizes of the two. This requires the electronic structure calculation to give not only the right result (gap value), but also the right result for the right reason (gap physics). In practice, a straightforward way of tackling the gap controversy is to treat the "true" gap of the antifluorite structure obtained under the correct gap physics as the contribution from the electronic mechanism. The difference between this and the experimental value is then used to roughly assess the magnitude of the structural contribution. Nevertheless, DFT-based methods such as aforementioned HSE and mBJ + $U$ both include tunable parameters. Based solely on the fact that the gap value is reproduced, it is not possible to determine whether this is a result of the tunable parameters or whether the gap physics is truly revealed. Therefore, the key to resolving the Cu$_{2}X$ gap controversy lies in identifying the calculation of correctly capturing the gap physics, which is what this work is addressing.

In this work, we find that the presence or absence of a computational gap in Cu$_{2}X$ depends on the competition between the 4$s$-3$d$ splitting of Cu atoms and the crystal field effect of anions. A severe underestimation of the former causes 3$d$ and 4$s$ band-order inversion and the Fermi energy pinning by symmetry, producing a zero gap. This is the reason why many DFT-based methods fail to open a gap. Using hybrid pseudopotentials (HYPPs) corrects the atomic splitting to the experimental value, which simultaneously opens a gap of the experimental magnitude in Cu$_{2}X$. As such, it not only gets the right results, but more importantly, it gets the right results for the right reasons. In contrast, neither Hubbard + $U$ nor hybrid functional calculations can correct for both atom-side and solid-side underestimations, implying an error cancellation. These results indicates a primarily electronic nature of the Cu$_{2}X$ gap.

\section{Methodology and models}

Three cross PP@functional calculations were performed using the Quantum Espresso package\cite{QE}, i.e., HYPPs\cite{YangJ} with PBE functional\cite{PhysRevLett.77.3865}, PBE PPs with PBE + $U$\cite{DFT+U} and HSE functional\cite{Heyd}, respectively denoted as HYPP@PBE, PBE@PBE$U$ and PBE@HSE below, for clarity. The HYPPs\cite{YangJ} were derived from the PBE0 functional, but with the amount of exact-exchange set as an adjustable parameter $\alpha$. Throughout this work, they involved only the Cu element and were constructed with the OPIUM code\cite{OPIUM} using the same parameters as before\cite{Wu, Shen, Bu, TanJCP}. For other elements, the norm-conserving Vanderbilt PPs\cite{PhysRevB.88.085117} were used with energy cutoffs of 60 Ry. A $12 \times 12 \times 12$ $k$-grid was used for the Brillouin zone sampling. Spin-orbit coupling has been neglected. While test calculations show that it does have an effect on the gap size, especially for Cu$_2$Te containing the heavy element, it does not affect the nature of the gap, which is the concern of this work. Optical absorption spectrum was obtained by solving the Bethe-Salpeter equation (BSE)\cite{PhysRevB.62.4927} using the YAMBO code\cite{marini2009yambo}, with three valence and three conduction bands to build the BSE Hamiltonian.

\section{Results and discussion}

Cu$_2$$X$ has similar physics except for the gap size, so Cu$_2$S is illustrated as an example below. Figure 1(a) shows its standard PBE band structure, which features a zero gap with the lowest conduction band and the two highest valence bands intersecting at the $\Gamma$ point. Fat-band analysis shows that this triple degeneracy originates from the higher-lying $t_{\rm 2g}$ triplet from the 3$d$-orbital splitting under the S tetrahedral field, while the Cu 4$s$ resides between it and the lower-lying $e_{\rm g}$ doublet. These results are consistent with previous studies\cite{s-lda-prb-2007,se-jpcm-2013,x-mbj-jcp-2014}. Despite a zero global gap, it is customary to define the $t_{\rm 2g}$-4$s$ difference at the $\Gamma$ point as a negative gap.

Two anomalies are worth mentioning about Fig. 1(a). On the one hand, it has been experimentally demonstrated that the oxidation state of Cu in Cu$_{2-x}$S is invariably +1 valence regardless of stoichiometry\cite{Coughlan}. Based on the unique 3$d^{10}$4$s^1$ electronic configuration of Cu, it is natural to speculate that Cu loses 4$s$ electrons to form a $d^{10}$ semiconductor when it crystallizes with S as Cu$_2$S. Consequently, 4$s$ and 3$d$ should comprise the conduction band bottom and the valence band top, respectively, instead of the zero gap and inverted 3$d$-4$s$ band-order in Fig. 1(a). On the other hand, one can clearly see that the band-order inversion occurs only very close to the $\Gamma$ point. A slight deviation will return the expected 4$s$-$t_{\rm 2g}$ order. Such discontinuity in state characteristics is physically uncommon.

\begin{figure}[htbp]
\begin{center}
\includegraphics[width=0.9\columnwidth]{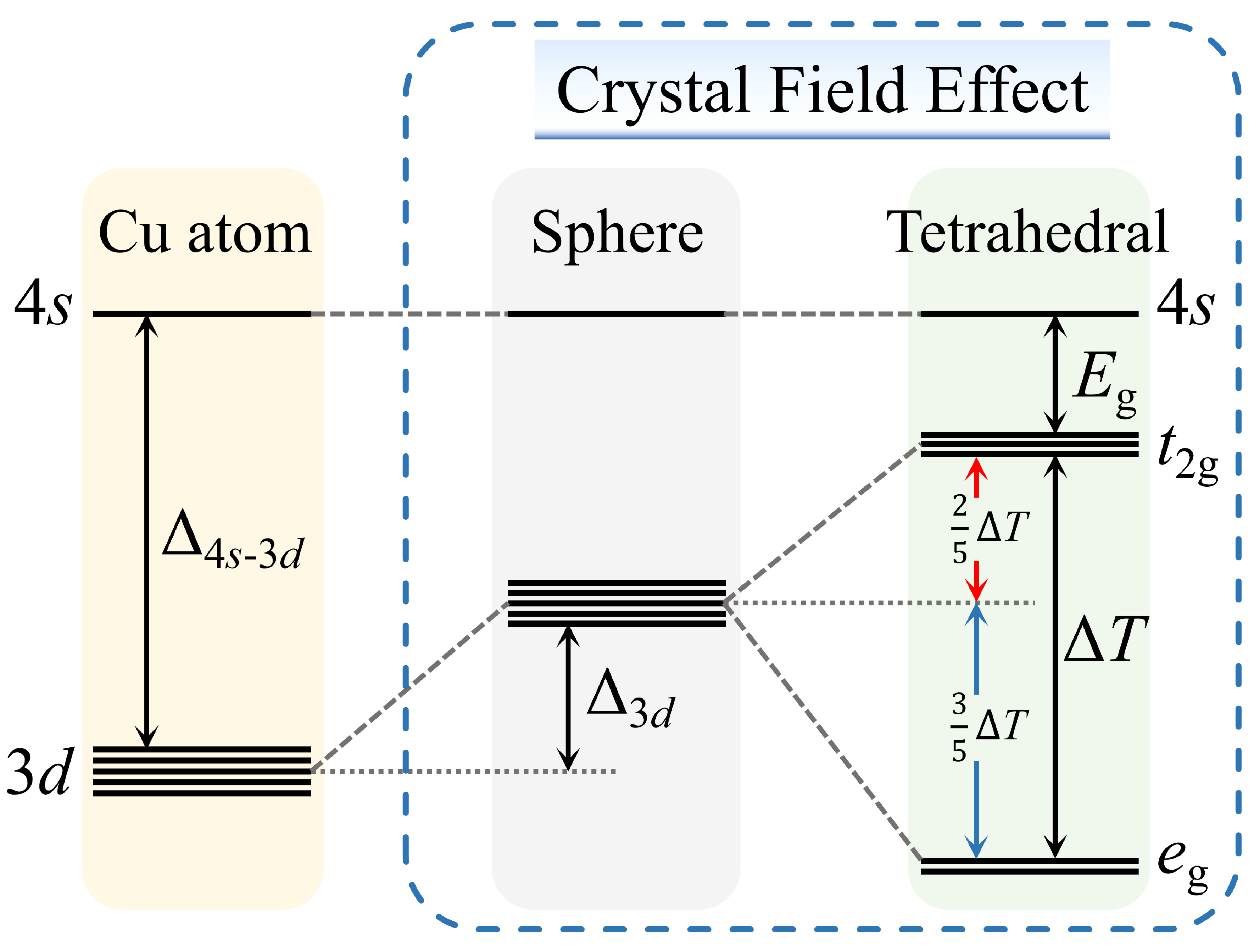}
\caption{\label{fig:fig2} Schematic of the Cu$_2X$ gap as a result of the competition between the 4$s$ and 3$d$ energy splitting $\Delta_{4s-3d}$ of Cu atom and the anionic crystal field effect. Upon crystallization into Cu$_2X$, the crystal field formed by the four surrounding anions narrows the splitting by $\Delta_{\rm {cf}}$ = $\Delta_{3d}$ + $\frac{2}{5}\Delta T$ in two ways. Assuming that the charge is uniformly distributed over a sphere, electrostatic repulsion causes the 3$d$ orbitals to rise $\Delta_{3d}$ as a whole. Consider further that the charges are distributed tetrahedrally. The 3$d$ orbitals are split into a higher-lying $t_{\rm 2g}$ triplet and lower-lying $e_{\rm g}$ doublet with an energy difference of $\Delta T$. If $\Delta_{4s-3d}$ surpasses the crystal field effect, $E_{\rm g}$ is positive. Otherwise, $t_{\rm 2g}$ will lie above the 4$s$, resulting in a negative $E_{\rm g}$.}
\end{center}
\end{figure}

Figures 1(b)-1(e) show the band structures by different Cu HYPPs of $\alpha$ = 0.25, 0.50, 0.75, and 1, in combination with PBE functional. One can see that Cu 4$s$ rises with increasing $\alpha$, from below $t_{\rm 2g}$ at $\alpha$ = 0 to coinciding with it at $\alpha$ = 0.5. Continuing to increase $\alpha$, a global gap appears and gradually increases. Once the gap opens, $t_{\rm 2g}$ and 4$s$ comprise the valence band top and the conduction band bottom, respectively. At $\alpha$ = 1, the gap reaches 1.20 eV, agreeing almost perfectly with the experimental 1.21 eV\cite{s-gap-1965}. Bader charge analysis shows that as $\alpha$ increases from 0 to 1, there is a small increase about 10\% in charge transfer from Cu to S.

We further calculate the optical absorption spectrum on top of the $\alpha$ = 1-HYPP@PBE band, as plotted in Fig. 1(f). Its first peak appears at 1.06 eV, which lies within the minimum direct gap. Therefore, its optical gap is set by an exciton with a binding energy of 0.14 eV.

It is clear that the band-order of $t_{\rm 2g}$ and 4$s$ determines the presence or absence of a gap in Cu$_2$S. A gap exists only for a positive 4$s$-$t_{\rm 2g}$ order. A negative 4$s$-$t_{\rm 2g}$ order, otherwise, causes the Fermi energy pinned to $t_{\rm 2g}$, resulting in an invariant zero gap at the $\Gamma$ point. Because the band-order depends on $\alpha$, an immediate question is which $\alpha$ is physically reasonable. It determines whether the experimental result reproduced at $\alpha$ = 1 comes from the right cause or is merely a coincidence of error cancellation.

The gap $E_{\rm g}$ of Cu$_2$S at the $\Gamma$ point can be understood in terms of crystal field theory, as illustrated in Fig. 2. An isolated Cu atom has an energy splitting $\Delta_{4s-3d}$ between 4$s$ and 3$d$ states. Upon crystallization into Cu$_2$S, the crystal field of S will narrow the splitting by $\Delta_{\rm {cf}}$ in two ways\cite{Bethe,Vleck1,Vleck2}: 1) Assuming that the charge is uniformly distributed on a sphere around the Cu, electrostatic repulsion leads to an overall increase in the 3$d$ energy by $\Delta_{3d}$; and 2) Further consideration of the tetrahedral charge distribution, the 3$d$ degeneracy is lifted to the $t_{\rm 2g}$ triplet and $e_{\rm g}$ doublet with an energy difference of $\Delta T$. According to crystal field theory, $t_{\rm 2g}$ is $\frac{2}{5}$$\Delta T$ higher than that of the original 3$d$\cite{Wolfram}. Therefore, we have
\begin{equation} \label{eq1}
E_{\rm g} = \Delta_{4s-3d} - \Delta_{\rm {cf}} = \Delta_{4s-3d} - (\Delta_{3d} + \frac{2}{5}\Delta T).
\end{equation}

Equation (1) actually provides a criterion for determining whether a computational method yields the true gap. If a method does capture the gap physics, it must be able to simultaneously give the correct $E_{\rm g}$, $\Delta_{4s-3d}$, and $\Delta_{\rm {cf}}$. While $\Delta_{\rm {cf}}$ cannot be obtained directly from experiment, both $\Delta_{4s-3d}$ and $E_{\rm g}$ can. If both are correct, it follows immediately from Equation (1) that $\Delta_{\rm {cf}}$ must also be correct. Otherwise, if only $E_{\rm g}$ matches the experiment but $\Delta_{4s-3d}$ does not, this seemingly correct $E_{\rm g}$ does not represent the true gap, and is merely a coincidence of error cancellation between $\Delta_{4s-3d}$ and $\Delta_{\rm {cf}}$.

This criterion makes physical sense. The Hohenberg-Kohn theorem states that the ground-state electron density of an interacting electron system is uniquely characterized by the external potential\cite{HK}. Changing the PP means changing the external potential felt by the Cu valence electrons. Here the $\Delta_{4s-3d}$ serves in a way as a test of the correctness of the external potential. If a method gets the correct $\Delta_{4s-3d}$ and $E_{\rm g}$ on both the atom side and the solid side, there is no doubt that the correct gap is obtained under the correct external potential. So it is true. Otherwise, only $E_{\rm g}$ being correct and $\Delta_{4s-3d}$ being incorrect implies that the gap is obtained under an incorrect external potential. Thus, it is an error-canceling coincidence and the true gap obtained under the correct external potential will no longer be $E_{\rm g}$.

The calculated $\Delta_{4s-3d}$, $\Delta_{3d}$, $\Delta T$ and $E_{\rm g}$ with different HYPPs are listed in Table I and compared with the experimental $\Delta_{4s-3d}$ and $E_{\rm g}$. It can be seen that $\Delta_{4s-3d}$ by standard PBE is only 0.63 eV, which is nearly an order of magnitude smaller than the experimental 5.04 eV\cite{one-electron}. It corresponds to the most negative $E_{\rm g}$ -0.91 eV. Both $\Delta_{4s-3d}$ and $E_{\rm g}$ increase monotonically with $\alpha$. Interestingly, when $\alpha$ = 1, $\Delta_{4s-3d}$ = 4.72 and $E_{\rm g}$ = 1.20 eV reproduce the experimental values (5.04 and 1.21 eV) almost simultaneously. Since $\Delta_{4s-3d}$ is still slightly underestimated, it makes sense and physical logic that $E_{\rm g}$ is also slightly underestimated. Now we know that the conventional DFT underestimation of the Cu$_2$S $E_{\rm g}$ is due to an underestimation of Cu $\Delta_{4s-3d}$, which incorrectly results in it being smaller than the 3$d$ energy rise induced by the S crystal field. This consequently reverses the band-order of $t_{\rm 2g}$ and 4$s$, producing a negative $E_{\rm g}$ and pinning the Fermi energy to the $t_{\rm 2g}$ by symmetry. In light of the criterion established above, the ability of using HYPPs to synchronously correct the underestimations of $\Delta_{4s-3d}$ on the atom side and $E_{\rm g}$ on the solid side indicates that the $\alpha$ = 1-HYPP@PBE indeed captures the gap physics. Therefore, the produced $E_{\rm g}$ = 1.20 eV is the right result for the right reason, representing the true gap of the antifluorite structure.

We also performed PBE@PBE$U$ and PBE@HSE calculations and the results are summarized in Table I for comparison. Both $\Delta_{4s-3d}$ and $E_{\rm g}$ show a tendency to increase with increasing $U$ or $\alpha$. Specifically, for the PBE@PBE$U$,  $\Delta_{4s-3d}$ = 3.77 eV at $U$ = 7 eV is underestimated by 1.27 eV, corresponding to a negative $E_{\rm g}$ of -0.14 eV. Increasing $U$ to 10 eV, $\Delta_{4s-3d}$ = 5.14 eV is already slightly higher than the experimental 5.04 eV, but opens only a tiny $E_{\rm g}$ of 0.09 eV. To open the experimental $E_{\rm g}$ $\sim$1 eV, $U$ would have to reach 45 eV, at which point $\Delta_{4s-3d}$ is four times larger than the experiment. The ineffectiveness of applying $U$ can be understood as a consequence of the rapid increase in Coulomb repulsion with increasing $U$, which leads to a severe overestimation of the S crystal field effect, thereby canceling out the vast majority of the $\Delta_{4s-3d}$ increase. The presence of such depletion causes $E_{\rm g}$ to grow extremely slowly with increasing $U$.

\begin{table}
\caption{The calculated $\Delta_{4s-3d}$, $\Delta_{3d}$, $\Delta T$ and $E_{\rm g}$ from standard PBE and different HYPPs (HYPP@PBE), PBE + $U$ (PBE@PBE$U$) and HSE (PBE@HSE) calculations, along with the known experimental values for comparisons. The value of $\alpha$ denotes the portion of exact exchange used for HYPPs or HSE functionals. All units are in eV.}
\setlength{\tabcolsep}{2.5mm}{
\renewcommand\arraystretch{1.2}
\begin{tabular}{l   r@{.}l   r@{.}l   r@{.}l  r@{.}l}
  \hline \hline
       {Methods}  &  \multicolumn{2}{c}{$\Delta_{4s-3d}$}   &  \multicolumn{2}{c}{$\Delta_{3d}$}   & \multicolumn{2}{c}{$\Delta T$}   &  \multicolumn{2}{c}{$E_{\rm g}$}  \\
     \hline
        Experiment               &   5&04\cite{one-electron}    &   \multicolumn{2}{c}{---}     &   \multicolumn{2}{c}{---}    &   1&21\cite{s-gap-1965}       \\
        PBE                      &   0&63   &  0&81   &   1&83   &   -0&91      \\
        $\alpha$=0.25-HYPP@PBE   &   1&59   &  0&94   &   2&17   &  -0&22   \\
        $\alpha$=0.50-HYPP@PBE   &   2&13   &  1&11   &   2&39   &  0&06   \\
        $\alpha$=0.75-HYPP@PBE   &   3&27   &  1&47   &   2&94   &  0&62   \\
        $\alpha$=1.00-HYPP@PBE   &   4&72   &  2&01   &   3&78   &  1&20  \\
        PBE@PBE$U$=7             &   3&77   &  2&49   &   3&54   &  -0&14  \\
        PBE@PBE$U$=10            &   5&14   &  3&27   &   4&46   &  0&09  \\
        PBE@PBE$U$=45            &   21&67  & 13&28   &  18&37   &  1&04  \\
        PBE@HSE-$\alpha$=0.25    &   2&65   &  1&24   &   2&47   &   0&42   \\
        PBE@HSE-$\alpha$=0.375   &   3&70   &  1&44   &   2&85   &  1&12   \\
        PBE@HSE-$\alpha$=0.50    &   4&77   &  1&69   &   3&57   &   1&65   \\
  \hline

  \hline
            \hline

 \end{tabular}}
\end{table}

Previous study\cite{se-jpcm-2013} has found that the PBE@HSE $E_{\rm g}$ for Cu$_2$Se is sensitive to the range separation parameter $\omega$, which is non-zero only when $\omega <$ 0.2 \AA$^{-1}$. Here, we fix $\omega $ = 0.106 \AA$^{-1}$ for HSE06\cite{Heyd} but change the exact exchange weight $\alpha$. It can be seen from Table I that the PBE@HSE at typical $\alpha$ = 0.25 gives $\Delta_{4s-3d}$ = 2.65 and $E_{\rm g}$ = 0.42 eV, respectively, both of which are significantly underestimated. At $\alpha$ = 0.375, $E_{\rm g}$ = 1.12 eV is close to the experimental 1.21 eV\cite{s-gap-1965}, but the corresponding $\Delta_{4s-3d}$ = 3.70 eV is still underestimated by 1.34 eV. At $\alpha$ = 0.50, $\Delta_{4s-3d}$ = 4.77 eV approaches the experimental 5.04 eV\cite{one-electron}, but $E_{\rm g}$ = 1.65 eV already exceeds the experimental 1.21 eV by nearly 40\%. Even taking excitons into account (borrowing the binding energy of 0.14 eV from above $\alpha$ = 1-HYPP@PBE calculation), the relative error is still 25\%. These point out that the PBE@HSE, unlike the $U$ effect, tends to underestimate the crystal field effect of S, and thus gives the largest $E_{\rm g}$ for the same $\Delta_{4s-3d}$.

Above results clearly show that neither PBE@PBE$U$ nor PBE@HSE can yield the correct $\Delta_{4s-3d}$ and $E_{\rm g}$ at the same time. Within typical parameter ranges ($U$ < 10 eV and $\alpha$ < 0.3), both methods tend to underestimate $\Delta_{4s-3d}$, which leads to an underestimation of $E_{\rm g}$. The PBE@PBE$U$ opening $E_{\rm g}$ by 0.09 eV at $U$=10 eV suggests that the Cu$_2$S gap is largely irrelevant to Mott physics. Even leaving aside the physical plausibility of the parameters, the gap calculated by the PBE@HSE for the antifluorite structure under the correct external potential is already significantly larger than the experimental value (see PBE@HSE-$\alpha$ = 0.50 in Table I). This suggests that the PBE@HSE is not suitable for resolving the gap controversy, since at this point it implies a clearly implausible negative structural contribution. Putting these together, neither PBE@PBE$U$ nor PBE@HSE correctly captures the gap physics of Cu$_2$S, so even if they happen to get the correct $E_{\rm g}$, it is not true but only the result of an error cancellation.

In order to fully understand the gap nature, we have also investigated the following three structural effects. First, the Cu atoms deviate from high-symmetry positions in the antifluorite structure. Despite the dependence on the direction and magnitude of the deviation, our test calculations with $\alpha$ = 1-HYPP@PBE show that this induces a gap correction of no more than 0.1 eV. This finding is consistent with a previous study\cite{s-lda-prb-2007}. Second, we have directly calculated the monoclinic, hexagonal, and cubic Cu$_2$S with the realistic structures using $\alpha$ = 1-HYPP@PBE and obtained gaps of 1.56, 1.43 and 1.30 eV, respectively. They are at the same level as the HSE\cite{Xu} and experimental results. Given that the electronic mechanism opens a gap of 1.20 eV, the structural mechanism is indeed of secondary importance. Third, we have evaluated the effect of copper stoichiometry in terms of cubic Cu$_{1.8}$S\cite{s-lda-prb-2007,Xu}. The standard PBE opens a gap of 0.42 eV while our $\alpha$ = 1-HYPP@PBE opens a gap of 1.68 eV. Note that it has an experimental gap of 1.75 eV\cite{Reijnen}. Compared to GW 1.06 eV\cite{s-lda-prb-2007} and HSE 1.20 eV\cite{Xu}, our result is clearly much better.

Performing $\alpha$ = 1-HYPP@PBE calculations on Cu$_2$Se and Cu$_2$Te open direct gaps at the $\Gamma$ point of 0.75 and 1.39 eV, respectively, compared to their experimental values of 1.23 eV\cite{se-gap-1966} and 1.08 eV\cite{te-gap-1990}. The former is underestimated by almost 40\%, signifying an important complementary role of Cu structural disorder. Indeed, an extremely strong liquid-like effect is experimentally found in Cu$_2$Se\cite{x-zt-2012}. In contrast, the relatively small overestimation for Cu$_2$Te comes from our neglect of the spin-orbit coupling\cite{x-mbj-jcp-2014}. Hence, its gap should originate from an electronic mechanism.

\section{Conclusions}
In summary, we compare the electronic structure of Cu$_2$$X$ calculated with HYPPs, Hubbard + $U$ and hybrid functionals, with a focus on the gap nature. While each can reproduce experimental gaps under specific parameters, only the $\alpha$ = 1-HYPP@PBE method complies with the right results for the right reasons. The other two suffer from significant error cancellations. Consequently, it is easy to understand that the same $\alpha$ = 1-HYPP also works well for other monovalent copper halides\cite{Wu} and delafossite oxides\cite{Shen}. Our results indicate that the time-averaged antifluorite model reflects the gap physics of Cu$_2$S and Cu$_2$Te well due to their predominantly electronic nature. However, for Cu$_2$Se, deviations from the antifluorite model and liquid-like structural disorder must be taken into account.

Traditionally, the PPs are used to reproduce all-electron results with higher efficiency, so those consistent with exchange correlations are customarily employed. In practice, there is another side of the coin for PPs as well, as their specific form also substantially affects the external potential where the valence electrons are located. This, however, has not yet been emphasized. The Hohenberg-Kohn theorem restricts the DFT to producing the correct ground-state properties only at the correct external potential. If calculations are performed at the wrong external potential, DFT fails in principle. Utilizing PPs to correct the external potential from the atomic side provides a new perspective on solving the persistent gap problem that has long plagued the DFT field.

\begin{acknowledgments}
This work was supported by the Ministry of Science and Technology of China (Grant Nos. 2023YFA1406400 and 2020YFA0308800) and the National Natural Science Foundation of China (Grant Nos. 12074034 and 11874089).
\end{acknowledgments}

\subsection*{AUTHOR DECLARATIONS}
\subsection*{Conflict of Interest}
The authors have no conflicts to disclose.

\subsection*{Author Contributions}
\textbf{Jiale Shen:} Formal analysis (equal); Investigation (equal); Software (equal); Validation (equal); Writing-original draft (equal). \textbf{Haitao Liu:} Formal analysis (equal); Funding acquisition (equal); Methodology (equal); Validation (equal). \textbf{Yuanchang Li:} Conceptualization (equal); Formal analysis (equal); Funding acquisition (equal); Methodology (equal); Project administration (equal); Resources (equal); Supervision (equal); Validation (equal); Writing-original draft (equal); Writing-review $\&$ editing (equal).

\section*{DATA AVAILABILITY}
The data that support the findings of this study are available from the corresponding author upon reasonable request.


\end{document}